\documentclass[]{aa}  

\usepackage{tikz,xcolor,graphicx}
\graphicspath{{images/}}

\usepackage{txfonts}
\usepackage{lipsum}
\usepackage{subcaption}
\usepackage{lscape}
\usepackage{placeins}
\usepackage{hyperref}
\hypersetup{hidelinks}
\usepackage{bm}

\definecolor{lime}{HTML}{A6CE39}
\DeclareRobustCommand{\orcidicon}{%
        \begin{tikzpicture}
        \draw[lime, fill=lime] (0,0)
        circle [radius=0.16]
        node[white] {{\fontfamily{qag}\selectfont \tiny ID}};
        \draw[white, fill=white] (-0.0625,0.095)
        circle [radius=0.007];
        \end{tikzpicture}
        \hspace{-2mm}
}
\newcommand{\orcidVP}{\href{https://orcid.org/0000-0002-3031-062X}{\orcidicon}}

\newcommand{\der}{\mathrm{d}}

\newcommand{\pardline}[2][]{\partial #1 / \partial #2}
\newcommand{\totd}[2][]{\frac{\der  #1}{\der  #2}}

\newcommand{\DF}{\mathcal{F}}
\newcommand{\Deff}{D_\mathrm{eff}}

\begin{document}

\title{The geometric origin of the thermal eccentricity law}
\subtitle{}

\author{
    Václav Pavlík\inst{\ref{asu},\ref{iu},}\thanks{\email{pavlik@asu.cas.cz}}\orcidVP
}
\authorrunning{V. Pavlík}

\institute{
    Astronomical Institute of the Czech Academy of Sciences, Bo\v{c}n\'i~II~1401, 141~00~Prague~4, Czech Republic \label{asu}
    \and Department of Astronomy, Indiana University, Swain Hall West, 727 E 3$^\text{rd}$ Street, Bloomington, IN 47405, USA \label{iu}
}

\date{Received 23 August 2026; accepted September 2026}

\abstract
{The distribution $f(e)=2e$ is commonly referred to as the `thermal eccentricity distribution' of Keplerian binaries. However, the result does not require thermal equilibrium, and its usual derivations obscure why the distribution is linear in eccentricity.}
{We seek a simple geometrical interpretation of this result and its relation to more general eccentricity distributions.}
{We consider bound Kepler orbits in a Euclidean space of dimension $D\geq2$, with a phase-space distribution depending only on energy, and derive the eccentricity distribution from the Liouville measure.}
{We obtain
$f_D(e)=(D-1)\,e\,(1-e^2)^{(D-3)/2}$,
so that, within this family, $D=3$ is the only case for which the distribution function is linear in $e$. This follows from the $D-1$ transverse momentum dimensions and the Kepler period degeneracy.
We further show that two known generalisations -- namely a power-law weighting of the normalised angular momentum and a constant velocity anisotropy -- are, in fact, two representations of the same one-parameter family of eccentricity distributions.}
{}

\keywords{%
binaries: general --
celestial mechanics --
methods: analytical --
stars: kinematics and dynamics --
gravitation
}

\maketitle
\nolinenumbers

\section{Introduction}

The eccentricity distribution
\begin{equation}
    \label{eq:thermal}
    f(e)=2e \,,
    \qquad 0 \leq e < 1 \,,
\end{equation}
is commonly referred to as the thermal eccentricity distribution of Keplerian binaries. It appears naturally in statistical treatments of binary populations and has played an important role in stellar dynamics and celestial mechanics since the early work of \citet{Jeans1919} and \citet{ambartsumian1937}. The law was later discussed in the context of stellar-dynamical encounters and binary evolution, for example by \citet{heggie_thesis, heggie75} and \citet{heg96}.
In dense star clusters, encounters continually exchange energy and angular momentum between a binary star and its neighbours, and a long-standing question was whether such interactions push a binary population towards a thermal eccentricity distribution; and, if so, whether the outcome is universal or bears the imprint of how the binaries formed and evolved \citep[see also, e.g.][]{geller+2019}.

The adjective `thermal' can somewhat obscure the origin of Eq.~\eqref{eq:thermal}. In particular, Ambartsumian already showed that a Maxwell--Boltzmann distribution is not required and that the same law follows when the phase-space distribution function depends only on orbital energy. This result therefore shifts the explanation of Eq.~\eqref{eq:thermal} away from thermodynamics and towards the way phase-space volume is distributed among bound Keplerian orbits. Why, then, does the corresponding phase-space measure yield an eccentricity probability density that is linear in $e$? Here, we show that the answer follows from the geometry of the bound Kepler problem.

\section{The generalised eccentricity law}
\label{sec:eccentricity}

We consider the relative motion of two bodies in the Kepler potential

\begin{equation}
    \Phi(r) = -\frac{\kappa}{r} \,,
    \qquad \kappa \equiv GM\mu > 0 \,,
\end{equation}
where $M$ is the total mass and $\mu$ is the reduced mass, $G$ is the Newtonian gravitational constant, and $r$ is the magnitude of the radius vector, $\vec{r}$.
The Hamiltonian of the relative motion is
\begin{equation}
    H(r,\vec{p}) = \frac{p^2}{2\mu} + \Phi(r) \,,
\end{equation}
where $\vec{p}$ is the momentum vector. We choose units such that the reduced mass is set to unity, so that $\kappa=GM$ and the masses of the two binary star components enter the dynamics only through $\kappa$. The motion is bound, with orbital energy $E<0$, and takes place in a Euclidean configuration space of integer dimension $D\geq2$.\footnote{The corresponding phase space has therefore dimension $2D$.}
We assume that the phase-space distribution function of the ensemble depends only on orbital energy,
\begin{equation}
    \label{eq:DF_energy}
    \DF=\DF(E) \,,
\end{equation}
and that the ensemble is sampled according to the Liouville measure, i.e.~the phase-space volume element
\begin{equation}
    \label{eq:Liouville_full}
    \der \Gamma = \der^D \vec{r} \, \der^D \vec{p} \,,
\end{equation}
whose key property is that the dynamics conserve it.
Thus, at fixed energy, no additional angular-momentum dependence is introduced.

At any position $\vec{r}$, the momentum, $\vec{p}$, can be decomposed into radial and transverse parts,
\begin{equation}
    \vec{p}
    =
    p_r \hat{\vec{r}}
    +
    \vec{p}_{\perp} \,,
\end{equation}
where $\hat{\vec{r}}$ is the unit vector in the radial direction.
There is one radial momentum component and $D-1$ transverse components. The volume of a shell in the $(D-1)$-dimensional transverse momentum space is, therefore, proportional to
\(
    p_{\perp}^{D-2} \, \der p_{\perp} \,.
\)
Similarly, the volume of a shell between
$r$ and $r + \der r$
in configuration space is proportional to
\(
    r^{D-1}\,\der r \,.
\)
After integrating over the orientations of
$\vec{r}$ and $\vec{p}_{\perp}$,
the relevant part of the Liouville measure, Eq.~\eqref{eq:Liouville_full}, is
\begin{equation}
    \label{eq:Liouville}
    \der \Gamma
    \propto
    r^{D-1} p_{\perp}^{D-2}
    \, \der r \, \der p_r \, \der p_{\perp} \,.
\end{equation}
The omitted angular factors depend on $D$, but not on angular momentum
or eccentricity, and hence do not affect the normalized distributions
derived below.

Let $J$ denote the magnitude of the orbital angular momentum per unit reduced mass (i.e.~the specific angular momentum). For a
central force,
\begin{equation}
    J = r p_{\perp} \,,
\end{equation}
so that, at fixed $r$,
\begin{equation}
    p_{\perp} = \frac{J}{r}
    \qquad\text{and}\qquad
    \der p_{\perp} = \frac{\der J}{r} \,.
\end{equation}
Substitution into Eq.~\eqref{eq:Liouville} gives
\begin{align}
    r^{D-1}p_{\perp}^{D-2}\,\der p_{\perp}
    &=
    r^{D-1}
    \left(\frac{J}{r}\right)^{D-2}
    \frac{\der J}{r}
    \nonumber\\
    &=
    J^{D-2} \, \der J \,.
\end{align}
The phase-space measure then becomes
\begin{equation}
    \label{eq:JD}
    \der \Gamma
    \propto
    J^{D-2}\,
    \der J\,\der r\,\der p_r \,.
\end{equation}
The factor $J^{D-2}$ is purely geometrical and counts the available transverse momentum states.

We now compare bound Kepler orbits with the same energy, $E<0$. Because of the assumption in Eq.~\eqref{eq:DF_energy}, $\DF$ is constant on the chosen energy surface and is absorbed into the normalisation. From Eq.~\eqref{eq:JD}, the number of phase-space points with angular momentum between $J$ and $J+\der J$ is
\begin{equation}
    \label{eq:dN_E}
    \der N_E
    \propto
    J^{D-2}\,\der J
    \int
    \delta\!\left[E-H(r,p_r,J)\right] \,
    \der r \, \der p_r \,,
\end{equation}
where the Hamiltonian is
\begin{equation}
    H(r,p_r,J)
    =
    \frac{p_r^2}{2}
    +
    \frac{J^2}{2r^2}
    -
    \frac{\kappa}{r} \,.
\end{equation}
The integral in Eq.~\eqref{eq:dN_E} covers the radial motion at fixed $E$ and $J$.
For every allowed radius, the energy equation has two solutions for $p_r$, corresponding to inward and outward motion. Since
$\pardline[H]{p_r}=\dot r$,
we have
\begin{align}
    \label{eq:Tradial}
    \int
    \delta(E-H) \,
    \der r \, \der p_r
    &=
    2\int_{r_{\rm p}}^{r_{\rm a}}
    \frac{\der r}{|\dot r|}
    \nonumber\\
    &=
    T_r(E,J) \,,
\end{align}
where $r_{\rm p}$ and $r_{\rm a}$ are the pericentre and apocentre, respectively, and $T_r(E,J)$ is the orbital period.\footnote{Strictly, $T_r$ is a radial period, i.e.~the time required for one full oscillation in $r$ (e.g.~between two pericentric passages). For a closed Kepler orbit, this coincides with the orbital period; for an apsidally precessing orbit, the two generally differ. We keep the subscript $r$ and the explicit $(E,J)$-dependence to emphasise that the quantity in Eq.~\eqref{eq:Tradial} is the general case.} Equation \eqref{eq:dN_E} therefore becomes
\begin{equation}
    \label{eq:dN_E_Tr}
    \der N_E
    \propto
    J^{D-2}T_r(E,J)\,\der J \,.
\end{equation}

The Kepler problem has a period degeneracy, that is, at fixed energy, the orbital period is independent of angular momentum,
\begin{equation}
    \label{eq:period}
    T_r(E,J)
    =
    2 \pi \kappa (-2E)^{-3/2}
    \equiv T(E) \,.
\end{equation}
The period factor in Eq.~\eqref{eq:dN_E_Tr} is therefore constant for all $J$ at the chosen energy and can be absorbed into the normalisation. Hence, Eq.~\eqref{eq:dN_E_Tr} becomes
\begin{equation}
    \label{eq:dN_E_kepler}
    \der N_E
    \propto
    J^{D-2}\,\der J \,.
\end{equation}

At fixed energy $E$, the semimajor axis, and hence the overall scale of the orbit, is fixed, while $J$ determines its eccentricity and therefore its shape. Let $J_{\rm c}(E)$ denote the maximum angular momentum allowed at that energy $E$, attained by the circular orbit, and define
\begin{equation}
    \label{eq:jJJc}
    j \equiv \frac{J}{J_{\rm c}(E)} \,,
    \qquad 0 \leq j \leq 1 \,.
\end{equation}
After normalisation, the probability element at fixed energy derived
from Eq.~\eqref{eq:dN_E_kepler} is
\begin{equation}
    \label{eq:prob_j}
    \der\mathcal{P}_E
    =
    (D-1) j^{D-2} \, \der j \,.
\end{equation}
The cumulative probability up to $j$ is therefore
\begin{equation}
    \label{eq:cdf_j}
    \mathcal{P}_E (j' \leq j)
    =
    \int_0^j
    (D-1) (j')^{D-2} \, \der j'
    =
    j^{D-1} \,.
\end{equation}
Thus, in $D$ dimensions, $j^{D-1}$ is uniformly distributed on $[0,1]$. Geometrically, this reflects the fact that, at fixed energy, the cumulative phase-space volume for angular momenta up to $J$ scales as $J^{D-1}$.

For a Kepler ellipse,
\begin{equation}
    \label{eq:j2_e2}
    J^2=J_{\rm c}^2(E)(1-e^2)
    \qquad\Rightarrow\qquad
    j^2=1-e^2 \,,
\end{equation}
and hence
\begin{equation}
    \label{eq:dj_de}
    \left|\totd[j]{e}\right|
    =
    \frac{e}{j} \,.
\end{equation}
Changing variables from $j$ to $e$ at fixed energy, the corresponding
eccentricity probability density is
\begin{equation}
    f_D(e)
    =
    (D-1) j^{D-2} \left|\totd[j]{e}\right| \,.
\end{equation}
Plugging in Eqs.~\eqref{eq:j2_e2} and \eqref{eq:dj_de}, we get the generalised form
\begin{equation}
    \label{eq:fe_final}
    f_D(e)
    =
    (D-1) \, e \,(1-e^2)^{(D-3)/2} \,,
    \qquad 0 \leq e < 1 \,.
\end{equation}
This probability density is independent of $E$ and therefore holds on every bound energy surface.

Two remarks clarify the status of this result.
First, the assumption in Eq.~\eqref{eq:DF_energy} states that $\DF$ is constant on each fixed-energy surface, i.e.~that bound orbits of equal energy are populated with equal weight, irrespective of angular momentum. Energy is the necessary argument: it is conserved and fixes the surface (and overall scale) on which the ensemble is defined. The fact that $E$ drops out of the normalised $f_D(e)$ in Eq.~\eqref{eq:fe_final} follows from the derivation.
Second, the measure in Eq.~\eqref{eq:Liouville_full} is geometric and coordinate-independent: it is invariant under canonical transformations, and for non-canonical ones, including the Jacobian leaves $\der N = \DF \der\Gamma$ unchanged. The eccentricity distribution is therefore the same in any coordinate system. We would obtain a different distribution only if the density were declared flat in some new variables without transforming the measure accordingly, thus defining a different statistical ensemble.

\section{Geometrical interpretation}

\subsection{The meaning of $D$}
\label{sec:meaning_D}

In Eq.~\eqref{eq:fe_final}, $D$ is the dimension of the Euclidean space in which the Kepler problem is defined. Individual orbits remain planar, because a central force confines the motion to the plane determined by the initial position and momentum. Changing $D$ therefore does not change the shape of a Kepler ellipse. Instead, it changes the number $D-1$ of momentum directions transverse to $\vec{r}$ and consequently the phase-space volume associated with a given angular momentum.

In the physical three-dimensional case ($D=3$), Eq.~\eqref{eq:cdf_j} says that $j^2$ is uniformly distributed and, from Eq.~\eqref{eq:j2_e2}, the same is true of $e^2$. Equation~\eqref{eq:fe_final} then becomes
\begin{equation}
    f_3(e)=2e \,,
\end{equation}
that is, exactly as in Eq.~\eqref{eq:thermal}.
The exponent of $(1-e^2)$ vanishes only for $D=3$. Hence, the familiar `thermal' law is the unique member of  Eq.~\eqref{eq:fe_final} which is linear in $e$.

The cases $D\neq3$ describe the potential $\Phi(r)$ embedded in a higher- or lower-dimensional Euclidean space. They should not be identified with Newtonian gravity in $D$ dimensions, for which the radial dependence of the potential is generally different.
For example, in $D=4$ the transverse momentum space is three-dimensional. Its shell volume scales as
$p_\perp^2\,\der p_\perp$\,,
which gives
$\der N_E\propto J^2\,\der J$
and
\begin{equation}
    f_4(e)=3e\sqrt{1-e^2} \,.
\end{equation}
Compared with the physical three-dimensional case, this gives more phase-space weight to large angular momenta and therefore to less eccentric orbits.

\subsection{A half-chord interpretation}
\label{sec:chord}

The physical three-dimensional case also admits an equivalent interpretation through the hidden
$\mathrm{SO}(4)$
symmetry of bound Kepler motion discussed by \citet{GS1990}. At fixed negative energy, let $\vec{j}$ be the normalised angular-momentum vector with magnitude $|\vec{j}|=j$, see Eq.~\eqref{eq:jJJc}, and let
\begin{equation}
    \label{eq:LRL}
    \vec{e}
    = \frac{\vec{p} \times (\vec{r} \times \vec{p})}{\kappa}
      - \hat{\vec{r}}
\end{equation}
be the Laplace--Runge--Lenz vector \citep[see, e.g.][]{goldstein1975, poisson_will}. The vector $\vec{e}$ is dimensionless, points from the focus towards the pericentre, and its magnitude is the eccentricity, $|\vec{e}|=e$. Both $\vec{j}$ and $\vec{e}$ are conserved along the orbit, and they are also perpendicular, that is, $\vec{j}\cdot\vec{e}=0$.
Combining this with Eq.~\eqref{eq:j2_e2}, we obtain the unit vectors
\begin{equation}
    \vec{n}_{\pm} \equiv \vec{j} \pm \vec{e} \,.
\end{equation}
Their separation is
\begin{equation}
    |\vec{n}_{+} - \vec{n}_{-}| = 2e \,,
\end{equation}
so the eccentricity is half the chord length joining the corresponding points between $\vec{n}_{+}$ and $\vec{n}_{-}$ on the unit sphere.

For the energy-only ensemble considered above,
$\vec{n}_{+}$ and $\vec{n}_{-}$
are independent isotropic directions\footnote{For a single orbit, $\vec{n}_{+}$ and $\vec{n}_{-}$ are jointly determined by the same $\vec{j}$ and $\vec{e}$; the independence holds over the whole ensemble.} \citep[see also][]{basha+2025}. If $\theta$ is the angle between them, the chord relation gives
\begin{equation}
    e^2 = \frac{1 - \cos\theta}{2} \,.
\end{equation}
Since $\cos\theta$ is uniformly distributed on $[-1,1]$, it follows that $e^2$ is uniformly distributed on $[0,1]$, and we recover Eq.~\eqref{eq:thermal}.

\section{Relation to previous eccentricity laws}
\label{sec:previous_results}

\subsection{Power-law weighting of angular momentum}
\label{sec:makarov}

\citet{makarov2025} generalised the Ambartsumian approach by allowing the phase-space density at fixed energy to depend on normalised angular momentum.\footnote{Makarov denotes $\sqrt{1-e^2}$ by $\beta$. To avoid confusion with the velocity-anisotropy parameter introduced later, we retain our notation $j$.}
The corresponding probability element in $j$ is
\begin{equation}
    \label{eq:makarov_prob_j}
    \der\mathcal{P}_{\alpha'}
    =
    (2-\alpha') j^{1-\alpha'} \, \der j \,,
    \qquad \alpha' < 2 \,,
\end{equation}
where $\alpha'$ is their power-law index controlling the angular-momentum weighting.

Changing variables from $j$ to $e$ using
Eqs.~\eqref{eq:j2_e2} and \eqref{eq:dj_de}, we get
\begin{equation}
    \label{eq:makarov_fe}
    f_{\alpha'}(e)
    =
    (2-\alpha')e(1-e^2)^{-\alpha'/2} \,.
\end{equation}
This reproduces Eq.~(13) of \citet{makarov2025}.
Comparison with Eqs.~\eqref{eq:prob_j} and \eqref{eq:fe_final} shows that Makarov's angular-momentum and eccentricity distributions are identical to the dimensional family under the effective identification
\begin{equation}
    \label{eq:Deff_alpha}
    \Deff \equiv 3-\alpha'
\end{equation}
(see the meaning of $\Deff$ in Sect.~\ref{sec:meaning_Deff}).
For $\alpha'=0$, the family reduces to the Jeans--Ambartsumian law, with $\Deff=3$.

\subsection{Constant velocity anisotropy}
\label{sec:anisotropy}

\citet{HattoriYoshii2010} derived a related eccentricity distribution for a spherical system with constant velocity anisotropy, $\beta$. Although their application concerned stellar orbits in a galactic potential, its point-mass limit reduces to the Kepler problem we considered here. Their distribution function has the form
\begin{equation}
    \label{eq:HY_DF}
    \DF(E,J)
    =
    g(E) \, J^{-2\beta} \,.
\end{equation}

In three dimensions, Eq.~\eqref{eq:dN_E_kepler} gives $\der N_E\propto J\,\der J$. The additional factor in Eq.~\eqref{eq:HY_DF} therefore leads to
\begin{equation}
    \der N_E
    \propto
    J^{1-2\beta} \, \der J \,.
\end{equation}
After normalisation, the probability element in $j$ is
\begin{equation}
    \label{eq:HY_prob_j}
    \der\mathcal{P}_{\beta}
    =
    2 (1-\beta) j^{1-2\beta} \, \der j \,,
    \qquad \beta<1 \,.
\end{equation}
Transforming to eccentricity gives
\begin{equation}
    \label{eq:HY_fe}
    f_{\beta}(e)
    =
    2 (1-\beta)\,e\,(1-e^2)^{-\beta} \,,
\end{equation}
which is the point-mass result of \citet{HattoriYoshii2010}.
This distribution has the same form as our Eq.~\eqref{eq:fe_final} under the effective identification
\begin{equation}
    \label{eq:Deff_beta}
    \Deff = 3 - 2\beta
\end{equation}
(see the meaning of $\Deff$ in Sect.~\ref{sec:meaning_Deff}).

\subsection{The meaning of $\Deff$}
\label{sec:meaning_Deff}

Unlike the literal dimension $D$ introduced in Sect.~\ref{sec:eccentricity}, $\Deff$ is not a spatial dimension. It is the formal value of $D$ which reproduces a given angular-momentum weighting in a physical three-dimensional system. This identification is possible because the normalised distribution derived in Eq.~\eqref{eq:fe_final} can be formally extended from integer $D\geq2$ to real $D>1$.\footnote{The geometrical derivation requires an integer
number of spatial dimensions, but the resulting distribution is an
explicit function of $D$ and Eq.~\eqref{eq:prob_j} defines a normalised probability distribution for any real $D>1$.}

Equating the two expressions for $\Deff$ in Eqs.~\eqref{eq:Deff_alpha} and \eqref{eq:Deff_beta} gives
\begin{equation}
    \label{eq:alpha_beta}
    \alpha' = 2\beta \,.
\end{equation}
Thus, the Makarov and Hattori--Yoshii families are two parameterisations of the same normalised angular-momentum and eccentricity distributions. This correspondence has a direct geometrical interpretation. Radial anisotropy, $\beta>0$, increases the relative weight of low-angular-momentum orbits and acts, at the level of the angular-momentum measure, as though fewer transverse momentum directions were available; hence $\Deff<3$. Tangential anisotropy, $\beta<0$, favours larger angular momenta and acts as though more transverse directions were available, giving $\Deff>3$. Through Eq.~\eqref{eq:alpha_beta}, the same interpretation applies to $\alpha'$.
We note, however, that this equivalence concerns the normalised distributions, not the underlying dynamics of the systems.

\section{Conclusions}
\label{sec:conclusions}

The usual description of $f(e)=2e$ as a `thermal' eccentricity law hides its simpler origin. For a Kepler ensemble whose phase-space distribution depends only on energy, the shape of the eccentricity distribution is fixed by geometry, and the single geometric input is the dimension of space, see Eq.~\eqref{eq:fe_final}. This dimension sets how many directions lie transverse to an orbit and therefore how much phase-space volume is apportioned to low- versus high-angular-momentum ellipses (i.e.~to eccentric versus round orbits). This weighting yields a linear eccentricity distribution if and only if there are three spatial dimensions. Hence, the `thermal' label merely signals that information about angular momentum is absent from the phase-space density.

This also explains why the known eccentricity laws, including the power-law result of \citet{makarov2025} and the constant-anisotropy result of \citet{HattoriYoshii2010}, are not independent. Both redistribute phase-space weight between small and large angular momenta within the same underlying geometry (see Sect.~\ref{sec:meaning_Deff}), so they emerge as two parametrisations of a single distribution family.

Our derivation fixes the form of the distribution for the idealised ensemble whose density depends only on energy, but it does not claim that real binaries attain this state. Whether encounters, formation, and evolution push a population towards this occupancy, or instead imprint a signature of their own, is a dynamical question. In this sense, any departure from the linear law signals that some process has re-introduced an angular-momentum dependence.

\begin{acknowledgements}
I am grateful to Steve Shore for inspiring conversations that motivated me to revisit this problem, to Douglas Heggie for persuading the University of Cambridge to make a digital copy of his PhD thesis available, and to Zdeněk Pavlík for lifelong encouragement of my work.
I~acknowledge support from the project RVO:67985815 at the Czech Academy of Sciences.
I~thank the anonymous referee for their comments that helped me improve the manuscript.
\end{acknowledgements}

\bibliographystyle{aa}
\bibliography{aa62705-26}

\end{document}